\documentclass[pra,aps,twocolumn,showpacs,epsfig]{revtex4}
\usepackage{color}
\usepackage{graphicx,floatflt,amssymb,epsfig,epsf,amsmath} 

\begin{document}

\title{Energy level statistics of interacting trapped bosons} 

\author{Barnali Chakrabarti $^{1,2}$, Anindya Biswas$^{3}$, V. K. B. Kota$^{4}$, Kamalika Roy$^{2}$, Sudip Kumar Haldar$^{2}$} 

\affiliation{
$^{1}$ Instituto de Fisica, Universidade de S\~ao Paulo, CP 66318, 05315-970, S\~ao Paulo, SP Brazil.\\
$^{2}$Department of Physics, Lady Brabourne College, P1/2 
Suhrawardi Avenue, Kolkata 700017, India.\\
$^{3}$Department of Physics, University of Calcutta, 
92 A.P.C. Road, Kolkata 700009, India. \\
$^{4}$ Physical Research Laboratory, Navarangpura, Ahmedabad 380009, India.}
\begin{abstract} 
It is an well established fact that statistical properties of energy level spectra are the most efficient tool to characterize 
nonintegrable quantum systems. Statistical behavior of different systems like, complex atoms, atomic nuclei, two-dimensional Hamiltonians, 
quantum billiards and non-interacting many bosons have been studied. The study of statistical properties and spectral 
fluctuation in the interacting many boson systems have developed a new interest in this direction. Specially we are interested in the 
weakly interacting trapped bosons in the context of Bose-Einstein condensation (BEC) as the energy spectrum shows a transition 
from the collective to single particle nature with the increase in the number of levels. However this has received less attention as it is believed that the system may 
exhibit Poisson like fluctuations due to the existence of external harmonic trap. Here we compute numerically the energy levels of the zero-temperature many-boson systems which 
are weakly interacting through the van der Waals potential and are in the $3D$ confined harmonic potential. We study the nearest neighbour 
spacing distribution and the spectral rigidity by unfolding the spectrum.
It is found that increase in number of energy levels for repulsive BEC induces a transition from a Wigner like form displaying level 
repulsion to Poisson distribution for $P(s)$. It does not follow the GOE prediction.
For repulsive interaction, the lower levels are correlated and manifest level repulsion.
For intermediate levels $P(s)$ shows mixed statistics which clearly signifies the existence of two energy scales
 : external trap and interatomic interaction, 
whereas for very high levels the trapping potential dominates, genarating Poisson distribution. Comparison with mean-field results for lower levels are also presented. For attractive BEC near the critical point we observe the Shnirelman like peak near $s=0$ which signifies the presence of large number of quasi-degenerate states.
\end{abstract}
\pacs{03.75.Hh, 31.15.Ja, 05.45.Mt, 05.45.Pq}
\keywords{Bose Einstein condensation, Potential harmonics, Quantum chaos, Shnirelman peak, spacing correlation, level spacing distribution }
\maketitle
\section{Introduction}
\hspace*{.5cm}
Although there is no precise definition of the quntum chaos, but the statistical properties of the energy level 
spectra is often used to characterize level fluctuation in quantum systems. It is an well established fact that for the classically 
integrable systems the energy levels are uncorrelated and nearest neighbor level spacing distribution follows Poisson statistics \cite{1}, 
whereas classically chaotic systems are associated with spectral fluctuations and strong level repulsion between energy levels is 
described in random matrix theory [2-4]. It follows GOE (Gaussian orthogonal ensemble)
or GUE (Gaussian unitary ensemble) of random matrices depending whether the Hamiltonian has time-reversal 
symmetry or not \cite{2,4}. The spectral properties of many different many fermion quantum systems like atoms and atomic nuclei and also quantum 
billiards have already been studied \cite{3,4,5,6,7,8,9,10,11}. $P(s)$ distribution of nuclear data ensemble agrees very well with GOE and in the atomic spectra the nearest-neighbor spacing distribution shows Wigner type. In addition, recently there are some studies of spectral properties of non-interacting many particle 
(fermions or bosons) systems \cite{12} and interacting boson systems \cite{10,11,13,14,15}. \\
\hspace*{.5cm}
In the present work we undertake to study the quantum mechanical spectra, the statistical behavior of weakly interacting many-boson 
systems with an external harmonic confinement. This study is specially interesting for several reasons. Firstly it directly corresponds 
to the Bose-Einstein condensation in dilute atomic vapor \cite{16,17}. Secondly due to the presence of external confinement the energy spectrum 
shows a transition from collective to single particle nature \cite{18,19}. Apparently it appears that the system will exhibit 
most common Poisson statistics of integrable systems as at the near zero temperature the interaction energy is small compared to 
the trap energy. However from the earlier studies of different statistics, thermodynamic and dynamic properties of the system it is an  
established fact that interatomic interaction plays an important role even in the weakly interacting Bose gas \cite{16,17,18,19}. Naturally it leads us to be more curious in the study of level 
spacing distribution. It seems to contradict the usual expectation based on RMT. This is a different type of system where two energy scales coexist. One is the external trap which is 
characterized by the trap energy $\hbar\omega$ ($\omega$ is the external trap frequency). The other one is the interatomic interaction which is characterized by $Na_{s}$; where $N$ is the number of bosons in the trap and the 
properties of zero-temperature BEC are essentially characterized by the $s$-wave scattering length $a_{s}$. Thus the study of 
spectral statistics of such a realistic system may provide exciting information on the level correlation and may disagree the universal hypothesis of Bohigas, Giannoni and Schmit \cite{2}. Due to existence of two energy scales, the system does neither 
obey the regular Poisson distribution nor the GOE for a strongly chaotic system. We observe new features from the following study. Quantum mechanical 
spectra undergoes a transition, as a function of number of energy levels. It has already been observed both experimentally and theoretically 
that low-lying levels are strongly influenced by interatomic interaction \cite{18,19}. These levels are highly correlated and we observe close to 
Wigner distribution. The intermediate levels show a mixed statistics which is the overlap of both Poisson and Wigner distribution which clearly 
signify the co-existence of two energy scales. Thus the choice in the number of levels has a great influence in the statistical properties.
For higher levels (much above the chemical potential) the energy spectra is strictly dominated by the harmonic confinement and energy levels are almost equidistant by the 
amount $\hbar \omega$, similar to the non interacting harmonic oscillator and it generates Poisson 
distribution. To the best of our knowledge there is neither any systematic calculations nor any rigorous derivation in this direction. 
Here we tackle the problem by solving the trapped many-body system by an {\it ab initio} but approximate many-body technique \cite{20,21,22}. 
As it is complicated many-body problem, it is hard to present analytic studies. However our numerical 
study is also important to investigate statistical behavior of such a realistic condensate.\\
\hspace*{.5cm}
The paper is organized as follows. Sec II deals with the many-body technique which basically calculates the many-body 
effective potential. Choice of interaction and detailed calculation of energy levels are presented in Sec III. Sec IV 
deals with several statistical tools and numerical results. Sec.V concludes the summary.
\section{Many-body technique}
\hspace*{.5cm}
We start with the Hamiltonian of a (N+1) trapped boson systems as \cite{20,21}
\begin{equation}
\label{many-body-Sch-eqn}
H = -\frac{\hbar^2}{2m} \sum^{N+1}_{i=1} \nabla^2_i + \sum^{N+1}_{i=1} V_{trap}(\vec{x}_i) + 
\sum^{N+1}_{i,j=1,j<i} V (\vec{x}_i-\vec{x}_j) \hspace{0.5cm}\cdot
\end{equation}
where $V_{trap}(\vec{x}_i)$ is the external trapping potential and  $V (\vec{x}_i-\vec{x}_j)$ is the two-body pair interaction. We use the standard Jacobi 
coordinates defined as $\vec{\zeta}_{i}$ = $\left(\frac{2i}{i+1}\right)^{\frac{1}{2}}\left[ \vec{x}_{i+1}- \frac{1}{i} 
\sum_{j=1}^{i} \vec{x}_{j}\right]$, $(i=1,2,....N)$ and the center of mass through $\vec{R}$ = $\frac{1}{N+1}\sum_{i=1}^{N+1}\vec{x}_{i}$. 
Then the relative motion of the atoms is described in terms of $N$ Jacobi vectors $(\vec{\zeta}_{1}, \cdot\cdot\cdot, \vec{\zeta}_{N})$ as \cite{20,23}
\begin{equation}
\left[-\frac{\hbar^{2}}{m}\sum_{i=1}^{N}\nabla_{\zeta_{i}}^{2}+V_{trap}+
V(\vec{\zeta}_1, \cdot\cdot\cdot, \vec{\zeta}_N)
-E \right]
\Psi(\vec{\zeta_{1}},\cdot\cdot\cdot,\vec{\zeta_{N}})=0. 
\end{equation} 
Hyperspherical harmonic expansion method (HHEM) is a convenient tool in many-body physics \cite{23}, where 
the expansion basis of the many-body wave function is the hyperspherical harmonics (HH). As HH basis keeps all
 possible correlations, its direct application to trapped bosons in the condensate which contains at least 
few thousand bosons, is an impossible task. Very recently we have adopted a technique called 
potential harmonic expansion method (PHEM) which keeps all possible two-body correlations together with 
realistic interatomic interaction \cite{20,22}. For the dilute Bose gas, the effect of two-body correlations are important and one 
can safely ignore the effect of all higher-body correlations. That is when two atoms interact the rest bosons are inert 
spectators and for zero temperature BEC this is a justified approximation.
Thus for the spinless bosons, we decompose $\Psi$ in two-body Faddeev components
\begin{equation}
\Psi = \sum_{ij>i}^{N+1} \psi_{ij} 
\end{equation}
Hence $\psi_{ij}$ is a function of two-body separation vector only, besides the global 
length (hyperradius, see below). 
$\psi_{ij}$ (symmetric under $P_{ij}$) satisfy the Schr\"odinger equation 
\begin{equation}
(T-E+V_{trap})\psi_{ij} = -V(\vec{x}_{ij})\sum_{k,l>k}\psi_{kl}, 
\end{equation}
$T$ being the total kinetic energy; operating $\sum_{i,j>i}$ on both sides of Eq. (4) we get back the original 
Schr\"odinger equation. The hyperradius is defined as $r$ = $ \sqrt{ \sum_{i=1}^{N}\zeta_{i}^{2}}$. The hyperradius and $(3N-1)$ hyperangles (denoted by $\Omega_N$) together constitute $3N$ hyperspherical variables. The choice of Jacobi coordinates is not fixed as the labeling of the particle indices is arbitrary.
We choose a particular set for the $(ij)$ interacting pair, called the $(ij)$-partition, by taking $\vec{r}_{ij}$ as $\vec{\zeta}_{N}$, and $(\vartheta, 
\varphi)$ are two spherical polar angles of the separation vector ${\vec{r}}_{ij}$.
The angle $\phi$ is defined through $r_{ij}$ = $r \cos \phi$. For the remaining $(N-1)$ Jacobi  coordinates we define the hyperradius for
the partition $(ij)$ as $\rho_{ij}= \left [ \sum_{k=1}^{N-1}\zeta_{k}^{2}\right] ^{\frac{1}{2}}$ such that $\rho_{ij}^{2} + r_{ij}^{2} 
= r^{2}$ and $\rho_{ij}= r \sin \phi$. With this choice, the hyperspherical coordinates are 
\begin{equation}
(r,\Omega_{N})= (r, \phi, \vartheta, \varphi, \Omega_{N-1})
\end{equation}
where $\Omega_{N-1}$ involves $(3N-4)$ variables: $2(N-1)$ polar 
angles associated with $(N-1)$ Jacobi vectors $\vec{\zeta}_{1}, \cdot\cdot\cdot, \vec{\zeta}_{N-1}$ and $(N-2)$ angles defining the relative lengths of these Jacobi 
vectors~[23]. Then the Laplacian in 
$3N$-dimensional space has the form 
\begin{equation}
\nabla^{2} \equiv \sum_{i=1}^{N} \nabla _{\zeta_{i}}^{2} = \frac{\partial^{2}}{\partial r^{2}}+ 
\frac{3N-1}{r} \frac{\partial}{\partial r}+ \frac{L^{2}(\Omega_{N})}{r^{2}}, 
\end{equation}
$L^{2}(\Omega_{N})$ is the grand orbital operator in $D$ = $3N$ dimensional space. Potential harmonics for the $(ij)$-partition 
are defined as the eigenfunctions of $L^{2}(\Omega_{N})$ corresponding to zero eigenvalue of $L^{2}(\Omega_{N-1})$ .
The corresponding eigenvalue equation satisfied by 
$L^{2}(\Omega_{N})$ is \cite{23}
\begin{equation}
\left[L^{2}(\Omega_{N}) + {\mathcal{L}}({\mathcal{L}}+D-2)\right] {\mathcal P}_{2K+l}^{l,m}(\Omega_{ij})=0, \hspace*{.5cm}
{\mathcal{L}}= 2K+l \cdot
\end{equation}
This new basis is a subset of the HH basis and is called as potential harmonics (PH) basis. It does not contain any function of the coordinate $\vec{\zeta}_{i}$ with 
$i<N$ and is given by \cite{23} 
\begin{equation}
{\mathcal P}_{2K+l}^{l,m} (\Omega_{(ij)}) =
Y_{lm}(\omega_{ij})\hspace*{.1cm} 
^{(N)} P_{2K+l}^{l,0}(\phi) {\mathcal Y}_{0}(D-3),
\end{equation}
where $Y_{lm}(\omega_{ij})$ is the spherical harmonics 
and $\omega_{ij}= (\vartheta, \varphi)$. The function 
$^{(N)}P_{2K+l}^{l,0}(\phi)$ 
is expressed in terms of Jacobi 
polynomials and ${\mathcal Y}_{0}(3N-3)$ is the HH of order zero in
the $(3N-3)$ dimensional space, spanned by $\{\vec{\zeta}_{1}, \cdots, 
\vec{\zeta}_{N-1}\}$ Jacobi vectors \cite{23}.   
Thus the contribution to the grand orbital quantum number comes only from the interacting pair and the $3N$ dimensional Schr\"odinger equation 
reduces effectively to a four dimensional equation. 
The relevant set of quantum numbers are  three -- orbital $l$, azimuthal $m$ and grand orbital $2K+l$ for any $N$. 
The full set of quantum numbers are
\begin{eqnarray}
l_{1} = l_{2} = \cdots=l_{N-1}=0,   & &  l_{N} = l\\
m_{1} = m_{2}=\cdots=m_{N-1}=0,  &&   m_{N} = m\\
n_{2} = n_{3}=\cdots=n_{N-1} = 0, &&  n_{N} = K .
\end{eqnarray}
We expand $(ij)$-Faddeev component, $\psi_{ij}$, in the complete set of PH basis appropriate for the $(ij)$ partition :
\begin{equation}
\psi_{ij}
=r^{-(\frac{3N-1}{2})}\sum_{K}{\mathcal P}_{2K+l}^{lm}
(\Omega_{N}^{(ij)})u_{K}^{l}(r) \hspace*{.3cm}\cdot
\end{equation}
which includes only two-body correlations. Taking projection of the Schr\"odinger equation on a particular PH, a set of coupled differential 
equations (CDE) is obtained \cite{20,21} 
\begin{eqnarray}
\Big[&-&\dfrac{\hbar^{2}}{m} \dfrac{d^{2}}{dr^{2}} + \dfrac{\hbar^{2}}{mr^{2}}
\{ {\cal L}({\cal L}+1) + 4K(K+\alpha+\beta+1)\} 
\nonumber 
\\
&-& E +V_{trap}(r) \Big] U_{Kl}(r) 
\\
&+& \sum_{K^{\prime}}f_{Kl}V_{KK^{\prime}}(r)f_{K'l} U_{K^{\prime}l}(r) = 0 ,
\nonumber
\end{eqnarray}
$U_{Kl}(r) = f_{Kl}u_{K}^{l}(r)$,  ${{\cal L}} =
l+\frac{3N-3}{2}$, $\alpha=\frac{3N-8}{2}$, $\beta=l+\frac{1}{2}$, $l$ being the orbital angular momentum contributed by the 
interacting pair and $K$ is the hyperangular momentum quantum number.
$f_{Kl}$ is a constant representing the overlap of the PH for interacting
partition with the full set, which is given in \cite{20}.  The
potential matrix element $V_{KK^{\prime}}(r)$ is given by \cite{20}
\begin{equation}
 V_{KK^{\prime}}(r) = \int {\mathcal P}_{2K+l}^{{lm}^{*}}(\Omega^{ij}_{N})
V(x_{ij})
{\mathcal P}_{2K^{\prime}+l}^{lm}(\Omega^{ij}_{N})d\Omega_{N}^{ij}.
\end{equation}
So far we have disregarded the effect of the strong short range correlation in the PH basis. 
In the experimentally BEC, the average interparticle separation is much larger than the range of two-body interaction. 
This is indeed required to prevent atomic three-body collisions and formation of molecules. As the energy of the 
interacting pair is extremely small, the two-body interaction is generally represented by the $s$-wave scattering 
length $(a_{s})$. A positive value of $a_s$ gives a repulsive condensate and a negative value of $a_s$ gives an 
attractive condensate. However a realistic interaction, like the van der Waals potential, is always associated with an 
attractive $-\frac{1}{x_{ij}^{6}}$ tail at larger separation and a strong repulsion at short separation. 
Depending on the nature of these two parts, $a_s$ can be either positive or negative. For a given finite range two-body potential 
$a_s$ can be obtained by solving the zero-energy two-body Schr\"odinger equation for the wave function $\eta(x_{ij})$
\begin{equation}
-\frac{\hbar^2}{m}\frac{1}{x_{ij}^2}\frac{d}{dx_{ij}}\left(x_{ij}^2
\frac{d\eta(x_{ij})}{dx_{ij}}\right)+V(x_{ij})\eta(x_{ij})=0  .
\end{equation}
The correlation function quickly attains its asymptotic form
$C(1-\frac{a_{s}}{x_{ij}})$ for large $x_{ij}$. The asymptotic
normalization is chosen to make the wavefunction positive at large
$x_{ij}$ \cite{22}. In the experimental BEC, the energy of the interacting pair is negligible 
compared with the depth of the interaction potential. Thus $\eta(x_{ij})$ is a good approximation of the short-range behavior of $\psi_{ij}$. 
Then we introduce this correlation function in the expansion basis and 
call it as correlated potential harmonics (CPH) basis. 
\begin{eqnarray}
\left[{\mathcal P}_{2K+l}^{l,m} (\Omega_{(ij)})\right]_{corr} =
Y_{lm}(\omega_{ij})\hspace*{.1cm} 
^{(N)} P_{2K+l}^{l,0}(\phi) \nonumber \\
 \times {\mathcal Y}_{0}(3N-3) \eta(x_{ij}) ,
\end{eqnarray}
$\eta(x_{ij})$ correctly reproduces the short separation behavior of the interacting-pair Faddeev component. Convergence rate of the 
PH expansion is quite fast. 
The
correlated potential matrix element $V_{KK^{\prime}}(r)$ is now given by
\begin{eqnarray}
V_{KK^{\prime}}(r) &=& (h_{K}^{\alpha\beta}
h_{K^{\prime}}^{\alpha\beta})^{-\frac{1}{2}}
\int_{-1}^{+1} \Big\{ 
P_{K}^{\alpha 
\beta}(z)
V\left(r\sqrt{\frac{1+z}{2}}\right) 
\nonumber 
\\
&&P_{K^{\prime}}^{\alpha \beta}(z)\eta\left(r\sqrt{\frac{1+z}{2}}\right)
w_{l}(z) \Big\} dz .
\end{eqnarray}
Here $h_{K}^{\alpha\beta}$ and $w_{l}(z)$ are respectively the norm
and weight function of the Jacobi polynomial
$P_{K}^{\alpha \beta}(z)$ \cite{20,21}. $K$ and $K^{\prime}$ are the grand orbital quantum numbers of the basis sets in which the potential matrix is calculated.
\section{ Choice of interaction and calculation of energy levels}
\hspace*{.5cm}
For the present study we consider few thousands (1000-10000) $^{87}$Rb atoms in the JILA trap \cite{16,17}. Throughout our calculation we choose 
$a_{ho} = \sqrt{\frac{\hbar}{m\omega}}$ as the unit of length (oscillator unit) and energy is expressed in units of the oscillator 
energy $(\hbar\omega)$. The van der Waals potential has been chosen as the interatomic potential with a hard core of 
radius $r_{c}$, \textit{viz}, $V(x_{ij})=\infty$ for $x_{ij} \leqslant r_c$ and $-\dfrac{C_6}{x_{ij}^6}$ for $x_{ij} > r_c$.
 The strength $C_6$ is taken as $6.4898\times10^{-11}$ o.u. for $^{87}$Rb atoms in the JILA experiment \cite{17}. We adjust the cut off 
radius $r_c$ in the two-body equation to correctly obtain $a_{s}=2.09\times10^{-4}$ o.u.\\
\hspace*{.5cm}
With these sets of parameters we solve the set of coupled differential equation [Eq. 13] by the hyperspherical adiabatic approximation (HAA) \cite{24}. 
We assume that the hyperadial motion is slow in comparison with the hyperangular motion. For the hyperangular motion for a fixed 
value of $r$, we diagonalize the potential matrix together with the hypercentrifugal term. Thus for a fixed value of $r$, the 
equation for the hyperangular motion can be solved adiabatically. The eigen value of this equation is a parametric function of $r$ and provides 
an effective potential for the hyperradial motion. In the HAA, the lowest eigenpotential is used for the ground state of the system and 
the hyperangular motion appears through the coupling matrix $V_{KK^{\prime}}(r)$. Thus the whole condensate collectively oscillates in the 
effective potential. The energy is thus obtained by solving the equation for the hyperradial motion as \cite{24}
\begin{equation}
\left[ -\frac{\hbar^{2}}{m}\frac{d^{2}}{dr^{2}} + \omega_{0}(r)+ \sum_{K=0}^{K_{max}} |\frac{d\chi_{K0}(r)}{dr}|^{2}-E \right] \zeta_{0}(r)=0 . 
\end{equation}
subject to the appropriate boundary conditions on $\zeta_{0}(r)$. This is called uncoupled adiabatic approximation (UAA), whereas disregarding the 
third term corresponds to extreme adiabatic approximation. HAA has already been successfully applied in different atomic and nuclear cases. \\
\hspace*{.5cm}
Although the lower multipolarities have been successfully detected in the experiments \cite{18}, however the collective 
excitations with higher multipolarity are also important specially to study the thermodynamic properties \cite{19}. In our many-body picture, the 
collective motion of the condensate is characterized by the effective potential as described earlier. Thus the excited states in this potential 
are the states with the $l$ th surface mode and $n$th radial excitation, which are denoted by $E_{nl}$. Thus $n$=0 and 
$l$=0 corresponds to the ground state and for $l \neq 0$, we get the surface modes. To calculate the higher levels with $l\neq 0$ we follow the next proedure. 
For $l$ $\neq$ 0, 
a large inaccuracy is involved in the calculation of off-diagonal potential matrix and numerical computation becomes very slow. However, the 
main contribution to the potential matrix comes from the diagonal hypercentrifugal term and we disregard the off-diagonal matrix element for 
$l>0$. Thus we get the effective potential $\omega_{l}(r)$ in the hyperradial space for $l\neq 0$. The energy of the lowest modes are in close 
agreement with the other calculations \cite{18,19,25,26} and we observe that for energy much larger than the chemical potential, the excited states are separated at 
energy close to the harmonic energy $\hbar\omega$ as in the noninteracting harmonic oscillator model. This transition from the 
low-energy collective to high-energy single particle spectrum leads us to study further the level fluctuation and other statistical behavior. \\
\hspace*{.5cm}
Before discussing the statistical behavior of the energy spectrum we discuss how good is our approximation method. There are many approximation methods 
to calculate the lowlying collective excitations and also the higher multipolarities. All these basically use the uncorrelated mean-field theory and the hydrodynamic method. Hydrodynamic method is good for large number of bosons in the trap in Thomas Fermi limit. Where as our system is finite sized and have few thousands bosons. As the system is not exactly solvable like the 1D system with contact $\delta$ interaction, it is not possible to calculate the accuracy of our approximation method. However from our previous calculation of different measurable quantities like crtical instability of attractive BEC, the collective excitations, the thermodynamic properties we observed that the   
correlated PHEM is an improvement over the Gross Pitaevskii mean field treatment for several reasons. Although the GP mean-field equation is being widely used, the wave function does not include any correlation. It is pointed out by several authors that the replacement of the actual interaction by a contact potential is not appropriate for a general realistic potential which consists of a repulsive core and an attractive part \cite{27,28,29}. The earlier studies also indicate the necessity of shape dependent potentials instead of the zero-range potential \cite{30,31}. The choice of contact 
interaction specially for $3D$ attractive BEC is not satisfactory \cite{30} as the $\delta$-function interaction produces an essential singularity at $r=0$. Thus to include correlations, one must go beyond the mean-field approximations and have to use the finite-range realistic potentials. Thus the correlated potential harmonics basis and the PHEM technique is a right step in this direction. This basis set retains all two-body correlations and assume that three 
and higher-body correlations are negligible. For dilute condensate it is perfectly justified. However including all two-body correlations we go beyond the mean-field theory. As the number of variables is reduced to only {\it{four}} for any number of bosons in the trap, we can treat quite a large number of atoms in the trap without much numerical complication. The use of van der Waals potential having a finite range takes care of the short-range repulsion and interatomic correlations. Clearly it is an improvement over the GP mean field theory. \\
\hspace*{.5cm} The correlated many-body aproach has been successfully applied in the calculation of static properties in different traps, collective excitations and thermodynamic properties of trapped bosons ~\cite{20,21,22,25,26,32,33,34,35}. In Refs \cite{25,26}, we have calculated the low-lying collective excitations by PHEM both for repulsive and attractive BEC. It has been shown that for repulsive and weak interaction the many-body results are close to the numerical solution of GP equation. However for large repulsive interaction significant difference is found. For the attractive BEC, the excitation frequencies for low lying modes are well comparable with the self-consistent Popov approximation. However higher multipolarities are lower than the GP result. This is attributed to the two-body correlations and finite-range interaction of the realistic interatomic interaction. In Ref \cite{32}, the 
PHEM is extended to investigate thermodynamic quantities which involves the calculation of a large number of energy levels. The calculated critical temperature and the condensate fraction have been calculated and compared with the GP results. The effect of realistic interatomic interactions and two-body correlations on thermodynamic properties of trapped bosons are observed. Thus the calculated energy levels are accurate for further analysis. We check for the convergence such that the error is considerable smaller that the mean level spacing.\\
\hspace*{.5cm}
Now to corroborate with the experiments we need the following discussions.
Exciting the condensate by applying inhomogeneous oscillatory force with tunable frequency, it is possible to observe several modes 
with different angular momentum and energy in the collective excitations ~\cite{18}. These experiments mainly concern low-lying 
collective excitations where the effect of interatomic interaction is prominant and the high-lying spectra should exhibit the 
single particle nature. The transition from collective to single particle excitations has also been studied in details theoretically ~\cite{19}. The collective 
excitations have also dramatic dependence on the temperature which comes from the interaction between the condensate and the thermal 
cloud ~\cite{36,37}. But for the present study we consider the zero temperature BEC and the effect of thermal fluctuations do not 
arise. There is no damping in the condensate as we assume there are no thermally excited atoms. Apparently it may contradict the 
experimental situation. But in the presence of external trapping and at zero temperature the effect of damping is not critical.\\
\section{ Statistical Tools and Numerical Results}
\hspace*{.5cm}
After getting the many-body collective levels including higher order excitations with different $l$, we transform the spectrum.
Next to characterize the spectral fluctuation in the many-body energy spectrum and to compare the statistical properties of different parts of the 
spectrum we remove the smooth part in the level density. In general, the level density has two parts : one is the smooth part which defines the 
general trend of the energy spectrum and a fluctuating part. The smooth part is removed by unfolding which maps the energy levels to another 
with the mean level density equal to 1. Several unfolding procedures are in the Ref ~\cite{4,38}. For the present calculation the many-body level 
density is approximated by a polynomial and unfolding is done by a 7th order polynomial. We unfold each spectrum separately for a specific value of $l$ and 
then form an ensemble having the same symmetry. Next in order to study the spectral fluctuation of this unfolded spectrum we utilize the 
following statistical tools. Nearest neighbor spacing distribution (NNSD) is the most applied tool in the study of the short range spectral correlations. 
From the unfolded spectrum, we calculate the nearest neighbor spacing as $s= E_{i+1}-E_{i}$ and calculate the probability distribution $P(s)$. 
Uncorrelated spectrum obeys the Poisson statistics and $P(s)=e^{-s}$. Whereas for the system with time-reversal symmetry, level 
repulsion leads to Wigner-Dyson distribution $P(s)=\frac{\pi}{2}s e^{-\frac{\pi s^2}{4}}$~\cite{39}. The $\Delta_{3}$ statistics is commonly 
used to investigate long-range correlations. It gives a statistical measure of the rigidity of a finite spectral level sequence. 
For a given interval $L$, it is often determined by the least square deviation of the staircase from the best straight line fits it.\\
\hspace*{.5cm}
The $P(s)$ distribution of the unfolded energy spectrum of 1000 interacting bosons in the external trap is presented in Fig.~1 (a)-(d)[see figure caption]. 
\vskip 0.5cm
\begin{figure}[hbpt]
\centerline{
\hspace{-3.3mm}
\rotatebox{0}{\epsfxsize=9cm\epsfbox{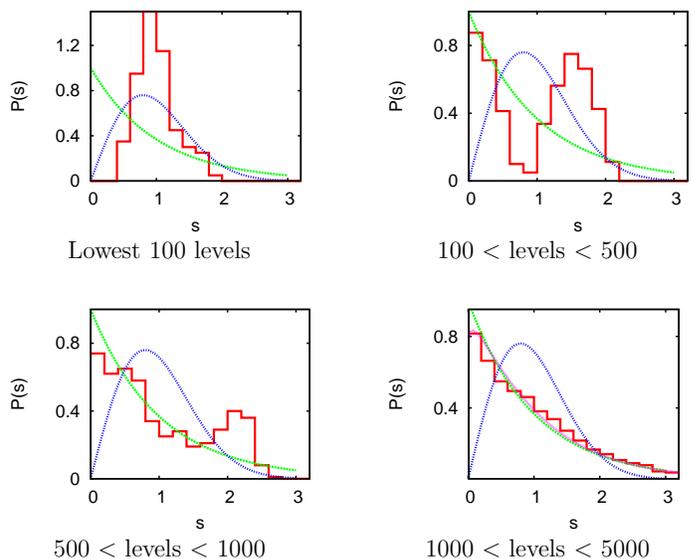}}}
\caption{(Color online) The $P(s)$ distribution is presented for different number of levels as indicated below the each panel, panel (a): lowest 100 levels, panel (b): 100 to 500 levels, panel (c): 500 to 1000 levels, panel (d): 1000 to 5000 levels. In each panel the histograms presents the $P(s)$ distrubution 
for the Hamiltonian (1) with $N=1000$ interacting bosons, the green dashed curve represents the Poisson distribution and the blue dotted curve represents the Wigner distribution. The magenta color solid line in the panel (d) corresponds to Brody distribution, corresponding Brody parameter being $\nu=0.04$.} 
\end{figure}

For comparison Poisson statistics and GOE statistics are also given in the same figure. For lowest 100 
levels there is no level with very small spacing and no level beyond $s=2.0$. Although the peak arises at $s \simeq 1.0$, it strongly 
deviates from Wigner distribution. For such low-lying levels, the effect of interatomic interaction is dominating and levels 
are highly correlated. This is also intuitively right as we may write the many-body effective potential in the following way  
\begin{equation}
\begin{array}{cl}
 \omega_{0}(r) & = V_{trap} + {\mathcal V}(r).\\
               & = \frac{1}{2} m\omega ^{2}r^{2} + {\mathcal V}(r)\\
\end{array} 
\end{equation} 
Where $V_{trap}$ is the external harmonic trap as described earlier and ${\mathcal V}(r)$ is obtained by the diagonalization of the potential matrix together with the hyper centrifugal repulsion.
Now for small value of $r$ (which corresponds to low-lying energy level) the effect of ${\mathcal V}(r)$ is dominating. Although it was expected that these levels would exhibit chaotic signature and follow Wigner distribution,
but the level repulsion is masked due to the existence of external harmonic trap. Thus it exhibits a mixed statistics which 
could not be perfectly interpolated between Poisson and Wigner distribution by using Brody parameter~\cite{9}. Thus the evolution 
of $P(s)$ distribution clearly shows the presence of two energy scales. The situation becomes more interesting for intermediate 
levels. The $P(s)$ distribution for 100 to 500 levels exhibits two peaks as shown in Fig.~1(b). The first narrow peak 
appears at $s=0$ with a second broad peak near $s=1.5$. For such intermediate levels, a part of the levels are correlated and 
shows normal level repulsion, whereas the other part do not repel each other and try to maintain Poisson statistics which is 
reflected as a first peak at $s=0$. The effect of level repulsion is manifested in the second peak. It is very similar to the 
classical mixed system, a part of phase space is completely regular with the other part chaotic. We have checked that by varying 
the number of levels in such an intermediate band, the width and peak values change but qualitative features remain the same. 
For much higher levels, the effect of interatomic interaction gradually decreases and the effect of harmonic trap starts to 
dominate. Thus more and more states are coupled in regular uncorrelated distribution. It is well reflected in Fig.~1(c),
where we see that a large part of levels try to exhibit Poisson statistics whereas a small fraction of levels is associated with
 a level repulsion, with a small peak near $s=2.0$. For much higher levels the effect of interatomic interaction is almost 
negligible and the levels become regular and close to the integrable system. $P(s)$ distribution is very close to Poisson, but 
the peak value at $s=0$ is less than 1. We fit the histogram with the Brody distribution [9],
\begin{equation}
\label{ps_fit}
 P(\nu, s)=(1+{\nu}) a s^{\nu} exp(-as^{1+{\nu}})
\end{equation}
where $a=\left[ \Gamma \left( \frac{2+{\nu}}{1+{\nu}} \right) \right]^{1+{\nu}}$. $\nu$ is the Brody parameter. The interesting feature of this 
distribution is that it interpolates between Poisson distribution $(\nu=0)$ of regular systems and the Wigner distribution with $\nu=1$. Thus 
the degree of chaos is determined by the value of $\nu$. For quantitative comparison, we fit the $P(s)$ histograms to $P(\nu,s)$ in Fig.~1(d) and 
the calculated 
repulsion parameter is ${\nu} = 0.04$. 
\vskip 0.5cm
\begin{figure}[hbpt]
\centerline{
\hspace{-3.3mm}
\rotatebox{0}{\epsfxsize=9cm\epsfbox{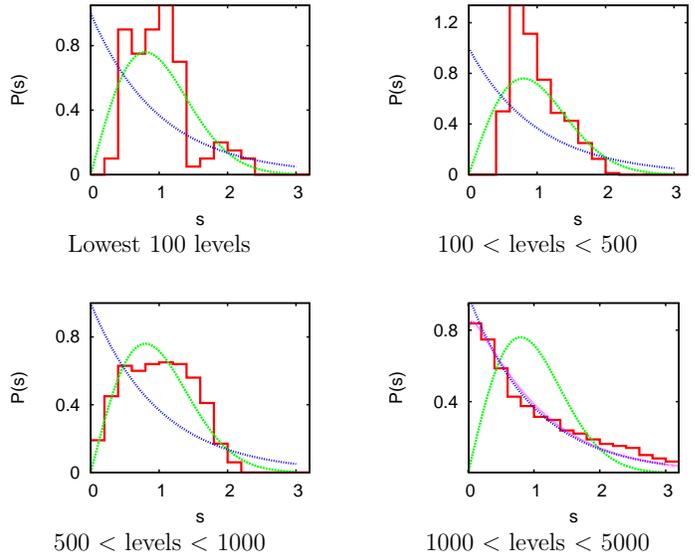}}}
\caption{(Color online) The $P(s)$ distribution is presented for different number of levels as indicated below the each panel, panel (a): lowest 100 levels, panel (b): 100 to 500 levels, panel (c): 500 to 1000 levels, panel (d): 1000 to 5000 levels. In each panel the histograms presents the $P(s)$ distrubution 
for the Hamiltonian (1) with $N=5000$ interacting bosons, the green dashed curve represents the Poisson distribution and the blue dotted curve represents the Wigner distribution. The magenta color solid line in the panel (d) corresponds to Brody distribution, corresponding Brody parameter being $\nu=0.05$.}
\end{figure}

The results for 5000 bosons is presented in Fig.~2(a)-(d)[see figure caption]. As the condensate is repulsive, with increase 
in particle number, the condensate wave function spreads out as the net effective repulsion $Na_{sc}$ increases. With increase in
 interaction more and more many-body levels show level repulsion and we expect level spacing distribution close to Wigner which is very 
similar to the completely chaotic system. But in our system as the level repulsion is suppressed by the external trap, the $P(s)$ distribution 
deviates from the Wigner distribution. This clearly shows the presence of two energy scales even for such intermediate levels. For 500
 to 1000 levels, we see quantum chaos sets in and $P(s)$ is very close to Wigner distribution. To observe and determine the 
best fit window to the Wigner, the $P(s)$ distribution for 501-600, 601-700 and 701-800 levels are plotted in Fig.~3. We
\begin{figure}[hbpt]
\centerline{
\hspace{-3.3mm}
\rotatebox{0}{\epsfxsize=9cm\epsfbox{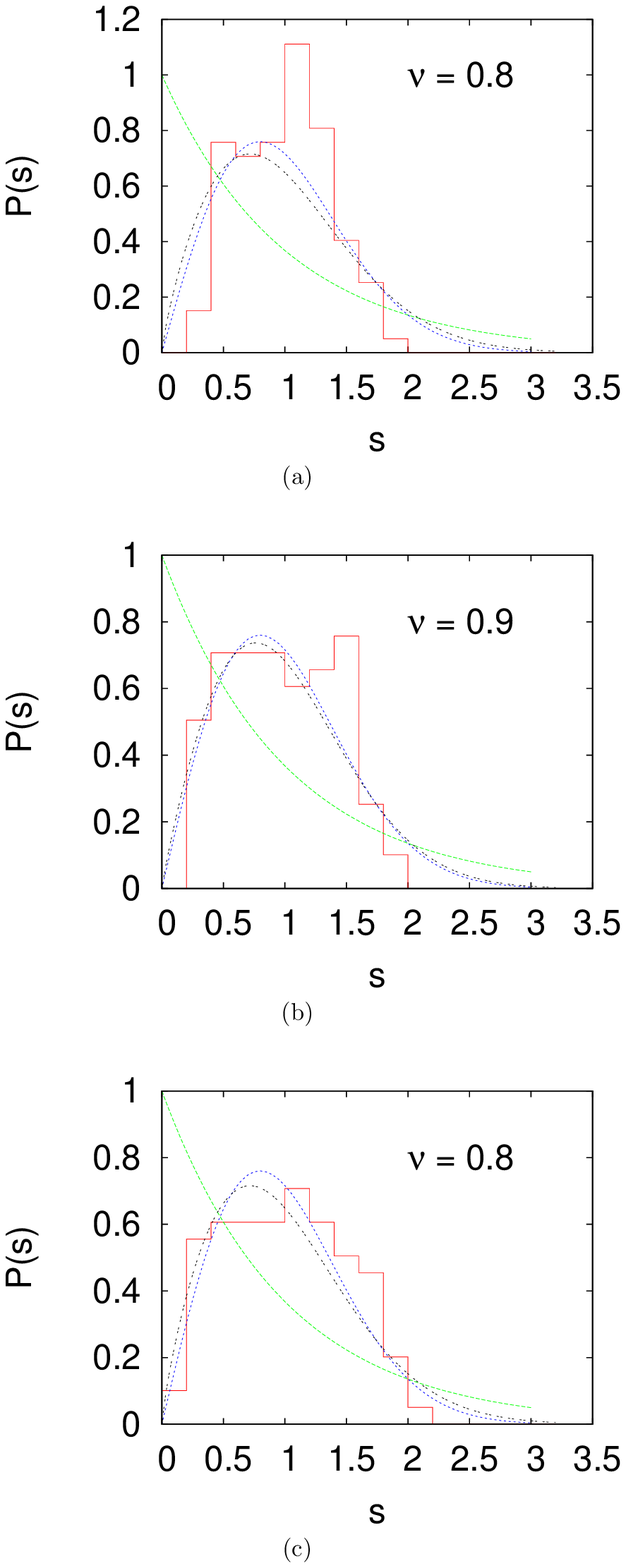}}}
\caption{(Color online) The $P(s)$ distribution is presented for different region of the spectrum, panel (a): 501-600 levels, panel (b): 601-700 levels, panel (c): 701-800 levels. In each panel the histograms presents the $P(s)$ distrubution 
for the Hamiltonian (1) with $N=5000$ interacting bosons, the green dashed curve represents the Poisson distribution, and the blue dotted curve represents the Wigner distribution. The black dot-dashed line corresponds to Brody distribution with the corresponding Brody parameter indicated in each panel.}
\end{figure}
observe that 601-700 is 
the best fit window and the corresponding Brody parameter is $\nu=0.9$. However this energy window strongly depends on the number of atoms and the scattering length, whereas for much higher levels, we
 observe a crossover from Wigner like level repulsion to Poisson. In Fig.~2(d) we again compare the histogram with the Brody distribution and 
the corresponding Brody parameter is ${\nu}= 0.05$.\\
\hspace*{1cm} 
We observe that $P(s)$ distribution depends crucially on the number of levels and on the net effective interaction $Na_{sc}$. To get more detail physical picture we calculate energy levels for 10000 bosons and plot the $P(s)$ distribution in Fig.4. Due to strong repulsive interaction $(Na_{sc}) \sim  2.09$, the low-lying levels are highly correlated and should show strong level repulsion. In Fig. 4, we present $P(s)$ distribution for the lowest 50 levels. Although the distribution has a sharp peak near $s$ = 0.8, the distribution shows Wigner-like level repulsion. Comparing the same for 5000 bosons [Fig.3], we observe the signature of level correlation for much lower levels. We also observe the uncorrelated Poisson distribution for higher levels as observed earlier.\\

\begin{figure}[hbpt]
\centerline{
\hspace{-3.3mm}
\rotatebox{0}{\epsfxsize=9cm\epsfbox{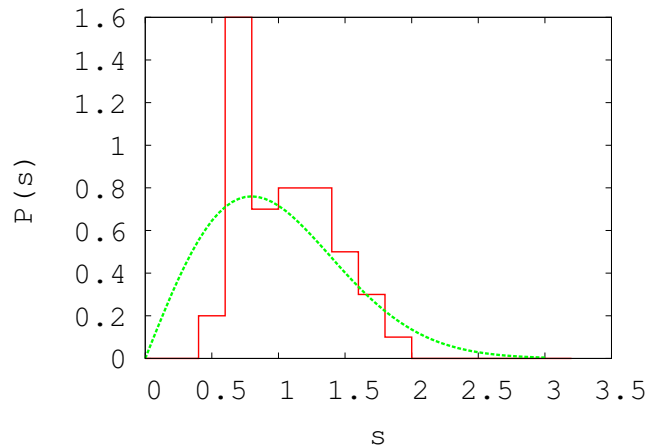}}}
\caption{(Color online) Plot of the $P(s)$ distribution for lowest 50 levels with 10,000 interacting bosons. Histograms represent the many-body result obtained for the Hamiltonian (1). Green dashed curve represents the Wigner distribution.} 
\end{figure}
\hspace*{1cm} 
Now in this connection it worths to calculate the spectral distribution for the energy levels which are calculated by Gross-Pitaveskii mean-field equation. As the mean-field equation uses the zero-range contact interaction and it completely ignores the interatomic correlation, it is interesting to observe the effect of finite range  interaction and interparticle correlation in the spectral statistics. For the calculation of energy levels we use the dispersion law of the discretized normal modes for spherical trap which is given by~\cite{n1}
\begin{equation}
 \omega(n_{r},l) = \omega {(2{n_{r}}^{2} + 2n_{r}l + 3n_{r} + l)}^{\frac{1}{2}} .
\end{equation}
$n_{r}$ is the radial quantum number and $n_{r}$=0 corresponds to surface excitation, whereas the monopole oscillation corresponds to $n_{r}$=1 and $l$=0. Note that Eq. (21) has the dependence only on the radial nodes and angular momentum, azimuthal degeneracy thus exists. Eq.(21) is also valid in the collisionless hydrodynamics where the number of atoms in the trap is quite high. For our chosen set, $N=5000$ and $a_{sc}$ = $2.09 \times 10^{-4}$ o.u., the parameter $\frac{Na_{sc}}{a_{ho}}$ $\simeq$ 1.045, which is just greater than one and Eq.(21) will be valid for lower $n_{r}$ and $l$. To see the accuracy of prediction [Eq.(21)], we plot in Fig.~5(a) the ground state energy per particle (E/N) as a function of $n_{r}$ for $l$=0. The GP results start to be lower and lower for larger $n_{r}$. The trend is maintained for other higher values of $l$. 
Our many-body results start to be higher near $n_{r}$ = 200 due to the presence of hyper centrifugal repulsion term in the many body equation [Eq. (13)]. As pointed in the Ref [40] that Eq. (21) is accurate for $\hbar\omega$ $<$ $\mu$. Thus for the calculation of $P(s)$ distribution using GP results we take lowest 500 levels for which the above condition is valid. It guarantees that our choice of levels will be reliable for the calculation of spectral distribution.
In Fig.~5(b) we 
\begin{figure}
  \begin{center}
    \begin{tabular}{cc}
      \resizebox{65mm}{!}{\includegraphics[angle=0]{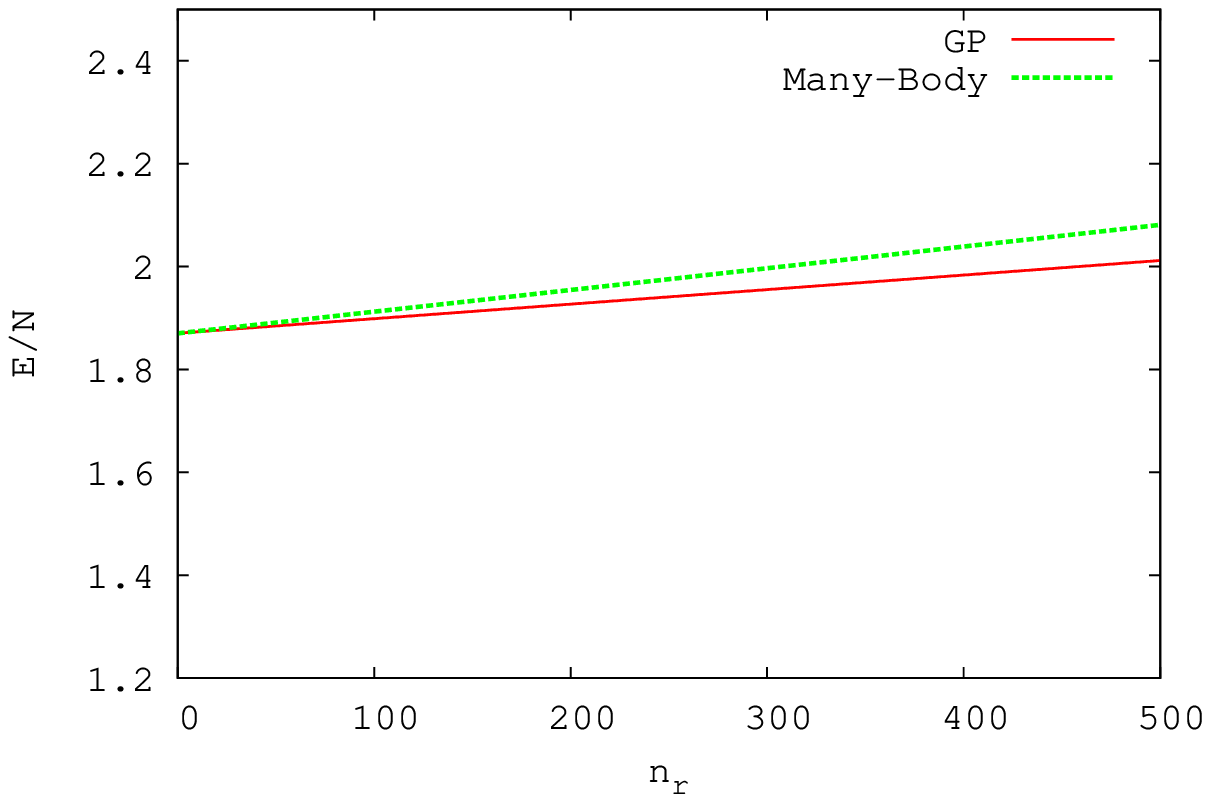}} \\
	(a)\\
      \resizebox{65mm}{!}{\includegraphics[angle=0]{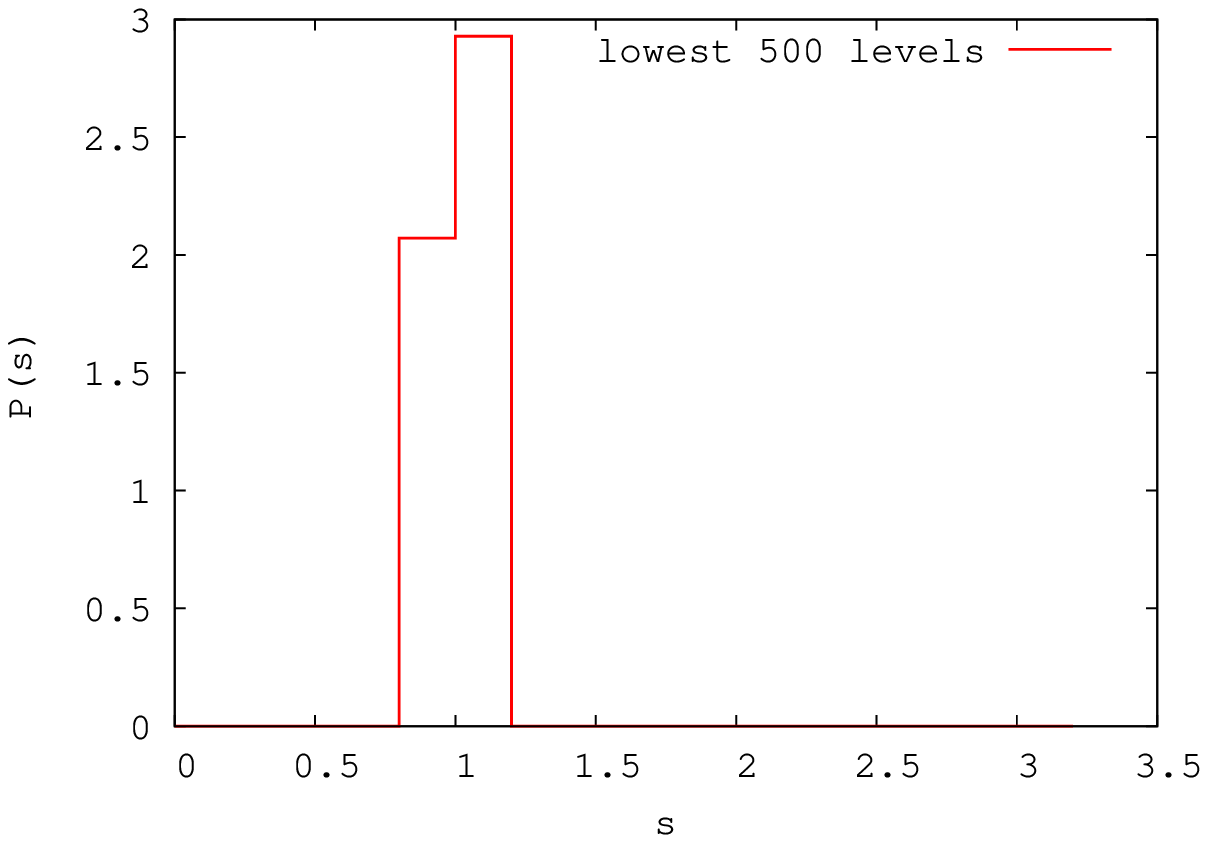}} \\
	(b)\\
    \end{tabular}
  \end{center}
\caption{(Color online) Fig.~5(a):  Plot of the ground state energy (in  o.u.) per atom (E/N), obtained from the many-body theory and GP equation,  as a function of $n_{r}$ for $l=0$. Fig.~5(b): The $P(s)$ distribution obtained from GP calculation is presented as Histogram.}
\end{figure}
plot the $P(s)$ distribution for lowest 500 levels obtained from the dispersion law.  It nicely shows the existence of large number of quasi-degenerate states as $P(s)$ exhibits a sharp peak near $s=1$. The existence of degeneracy is also seen in Eq.(21) where the discrete eigen mode frequency $\omega(n_r,l)$ of the spatial variation of density, obtained in the context of the hydrodynamic model of the condensate at low temperature, is a function of the radial quantum no $n_r$ and orbital quantum no $l$ only and hence is degenerate with respect to the azimuthal quantum no $m$. The contact $\delta$-potential in the GP equation can not lift this degeneracy. There is a $\delta$-type peak at about $s=1$.
 
Such $\delta$-type peak in the $P(s)$ distribution is called as Shnirelman peak~\cite{dpk,Chirikov}. In the year 1993, Shnirelman showed that systems with time reversal symmetry should exhibit the Shnirelman peak in the $P(s)$ distribution. 
This behavior is again expected from Eq.~(19). In the mean field results ${\mathcal V}(r)$ is calculated only taking the contact $\delta$ interaction and ignores the interatomic correlation completely. Thus it can not lift the degeneracy of the external harmonic trap completely. Whereas in our many-body calculation the short-range hard sphere below the cutoff radius and the -$\frac{C_6}{x^6}$ tail in the interatomic interaction takes care of the effect of both short-range and long-range correlation and gives actual physical picture. But $P(s)$ contains no information about spatial correlation. A simple measure of spacing correlation is the correlation coefficient $C$ defind as 
\begin{equation}
C= \sum_{i} (s_{i}-1)(s_{i+1}-1)/\sum_{i} (s_{i}-1)^{2}
 \hspace*{.3cm}\cdot
\end{equation}
For uncorrelated spectra $C=0$. The calculated value of $C$ for Fig.~5(b) is 1.0 which again signifies the existence of bulk quasi-degenerate states.\\
\hspace*{1cm}
Here we observe that $P(s)$ distribution strongly depends on the number of energy levels and also on the net effective interatomic interaction. 
Thus it is hard to say about correlation properties only from the study of $P(s)$ distribution. It represents the study of the correlation properties in the large energy scale which will give new physical insight. The spectral rigidity 
is often used as a stronger tool than the level distribution in the analysis of complex systems as it can take into account of the long-range 
correlation between the levels while $P(s)$ distribution takes into account only nearest neighbour correlations. So further studies 
are needed in this direction. We are mainly interested in  $\Delta_{3}$ statistics of Dyson and Mehata ~\cite{40} which gives a statistical 
measure of the rigidity of a finite spectral level sequence. For a level sequence with a constant average level spacing, the staircase function on the 
average follows a straight line. Thus $\Delta_{3}$- statistics gives a measure of the size of fluctuations of the staircase function about a best fit straight line.  For Poisson spectrum, the levels are 
uncorrelated, the spectrum is rigid and $\left\langle \Delta_3(L) \right\rangle = \frac{L}{15}$, whereas for GOE distribution,
levels are strongly correlated and $\left\langle \Delta_3(L) \right\rangle \propto log~L$. So to confirm our earlier findings in
 $P(s)$ distribution we next study the correlation structure and $\Delta_3$ statistics for our system which are shown in Fig.~6 and Fig.~7.
The results $<\Delta_{3}(L)>$ are plotted against $L$ for the same parameters and same number of levels as chosen in Fig.~1 and Fig.~2. Results for 1000 interacting bosons are presented in Fig.~6. For 
comparison we also plot $<\Delta_{3}(L)>$ for Poisson and GOE. For low-energy levels we observe the same trend, bending towards the GOE prediction.
\begin{figure}[hbpt]
\centerline{
\hspace{-3.3mm}
\rotatebox{270}{\epsfxsize=6cm\epsfbox{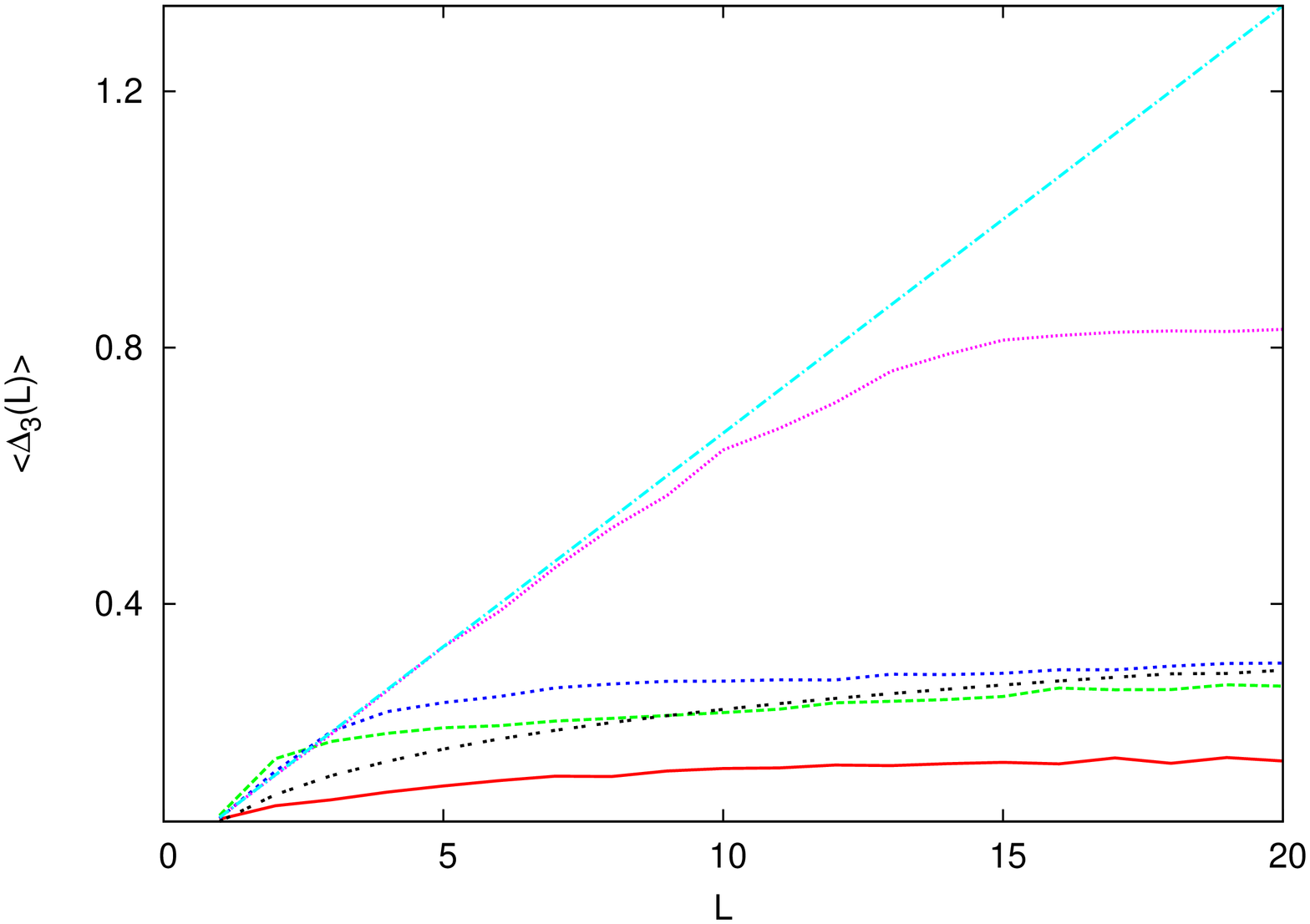}}}
\caption{(color onlne)
Spectral average $<\Delta_{3}(L)>$ computed for the Hamiltonian (1) with 1000 interacting bosons 
in the external trap. Red color corresponds to lowest 100 levels. Green color corresponds to levels between 100 and 500. 
Blue color corresponds to levels between 500 and 1000. Violet color corresponds to levels between 1000 
and 5000. The straight line corresponds to Poisson distribution and the black color corresponds to GOE result.
}
\end{figure}
However the saturated value is well below the GOE prediction. It again reflects the fact that level repulsion due to interatomic interaction 
is screened due to the effect of external harmonic trap. For 1000-5000 levels, $<\Delta_{3}(L)>$ follows the expected straight line behaviour 
upto $L\leq 10$, which is the result of integrable systems. But beyond $L=10$, it still tends to saturation, but it is consistent with Berry's semiclassical arguments ~\cite{41,42}.
\begin{figure}[hbpt]
\centerline{
\hspace{-3.3mm}
\rotatebox{270}{\epsfxsize=6cm\epsfbox{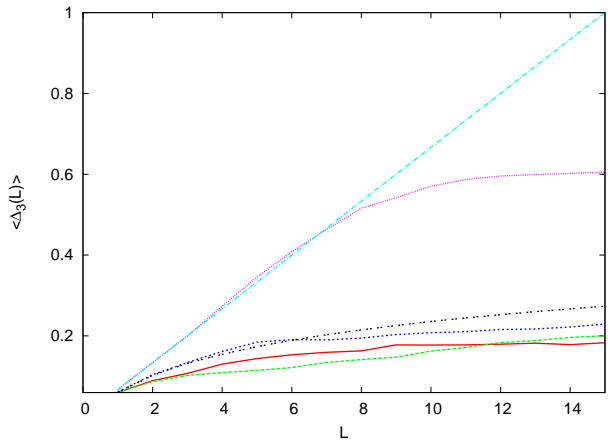}}}
\caption{(Color online) 
Spectral average $<\Delta_{3}(L)>$ computed for the Hamiltonian (1) with 5000 interacting 
bosons in the external trap. Red color corresponds to lowest 100 levels. Green color corresponds to levels 
between 100 and 500.
Blue color corresponds to levels between 500 and 1000. Violet color corresponds to levels between 1000
and 5000. The straight line corresponds to Poisson distribution and the black color corresponds to GOE result.
}
\end{figure}
The results for 5000 bosons are shown in the Fig.~7, which again indicates that lower levels are highly 
correlated whereas for higher levels, we get the signature of more regular distribution. \\
\hspace*{1cm}
It is useful to mention that Guhr and Weidenm\"uller~\cite{43} studied in the past the spectral properties of a regular Hamiltonain perturbed by  a GOE. 
The results for 100 levels and 100 to 500 levels shown in the present work are quite similar to some of the results in Figs.~1,2,3 and 6 
of~\cite{43} where a modified uniform spectrum was used as the regular Hamiltonian. Therefore, a quantitative description of the results in Figs.~1-5 in terms 
of a deformed GOE, which combines uniform, GOE and Poisson is possible but this is for a future investigation. This variety of behavior has also been 
observed early in the study of quantum mechanics of heavier clusters like $Kr$ and $Xe$ trimers. The energy spectrum of these clusters show a wide 
variety of behavior below and above the transition energy ~\cite{44}. Very regular behavior of low-lying eigenstates changes to the combination 
of regular and irregular behavior at energy above the transition energy.\\ 
\hspace*{1cm}
Experiments on Bose-Einstein condensation with $^{7}$Li atoms is another challenging research area where the $s$-wave scattering length ($a_{s}$) is negative, which indicates that atom-atom interaction is negative~\cite{n2}. A homogeneous condensate with a negative scattering length is impossible as the condensate approaches collapse. However the situation changes drastically in the presence of an external confinement. Spatially confined BEC is stable for a small, finite number of atoms ($N_{cr}$). For $^{7}$Li, $a_{s}$ = -27.3 Bohr = -45.7$\times{10}^{-5}$ o.u. and for $T=0$ a metastable condensate exists when the number of atoms is less than the critical number $\simeq$ 1300~\cite{n3}, whereas theory  predicts that BEC can occur in a trap with no more than about 1400 atoms \cite{n4}.\\
\hspace*{1cm}
    It is pointed out earlier in different connection~\cite{n5,n6} that the GP theory based on the pseudo potential form of the interatomic interaction is not suitable as an exact potential in $3D$ attractive system. Again as the attractive BEC becomes highly correlated near the critical points, the uncorrelated GP equation cannot take care of the effect of interatomic correlation. In our earlier calculation we have extensively applied our many-body method in the study of different properties and stability of the attractive condensate~\cite{n7,n8}. In our present study we are interested in the spectral distribution of highly correlated BEC. The presence of hard sphere below some cutoff radius and the -$\frac{C_{6}}{x^{6}}$ tail in the interatomic interaction properly takes care of the effect of both short-range and long-range correlations.\\
\hspace*{1cm}
 For N $<$ $N_{cr}$, the condensate is metastable. In the many-body effective potential, the intermediate metastable region (MSR) is bounded by the high wall of the external trap on the right side and a very deep narrow attractive well appears on the left side of left intermediate barrier~\cite{n8}. As the very high-lying levels will have large probability to tunnel through the barrier we are interested only for the low-lying levels for which the transition probability is almost zero. The results for $N=1000$ and $N=1300$ are presented in Fig.~8 and in Fig.~9 respectively. We do not get any stable solution beyond $N=1320$. So for our present calculation the critical number is $N_{cr}$=1320. For $N=1000$, as the number of atoms is well below the critical number, we correctly obtain the lowest 100 levels which are well bound within the metastable region. We observed high level correlation for lowest 100 levels and high-lying levels are uncorrelated. The effect of interatomic correlation for attractive BEC is very important for low-lying levels as the effect of ${\mathcal V}(r)$ dominates for smaller values of $r$. Although the level-correlation strongly dominates, however we do not observe any $\delta$ function like peak. It signifies that for $N=1000$, the exact degeneracy of harmonic oscillator is completely removed by the interatomic interaction ${\mathcal V}(r)$, whereas for larger $r$, as the term $\frac{1}{2} m\omega ^{2}r^{2}$ dominates we get the uncorrelated Poisson distribution. 
\begin{figure}
  \begin{center}
    \begin{tabular}{cc}
      \resizebox{65mm}{!}{\includegraphics[angle=0]{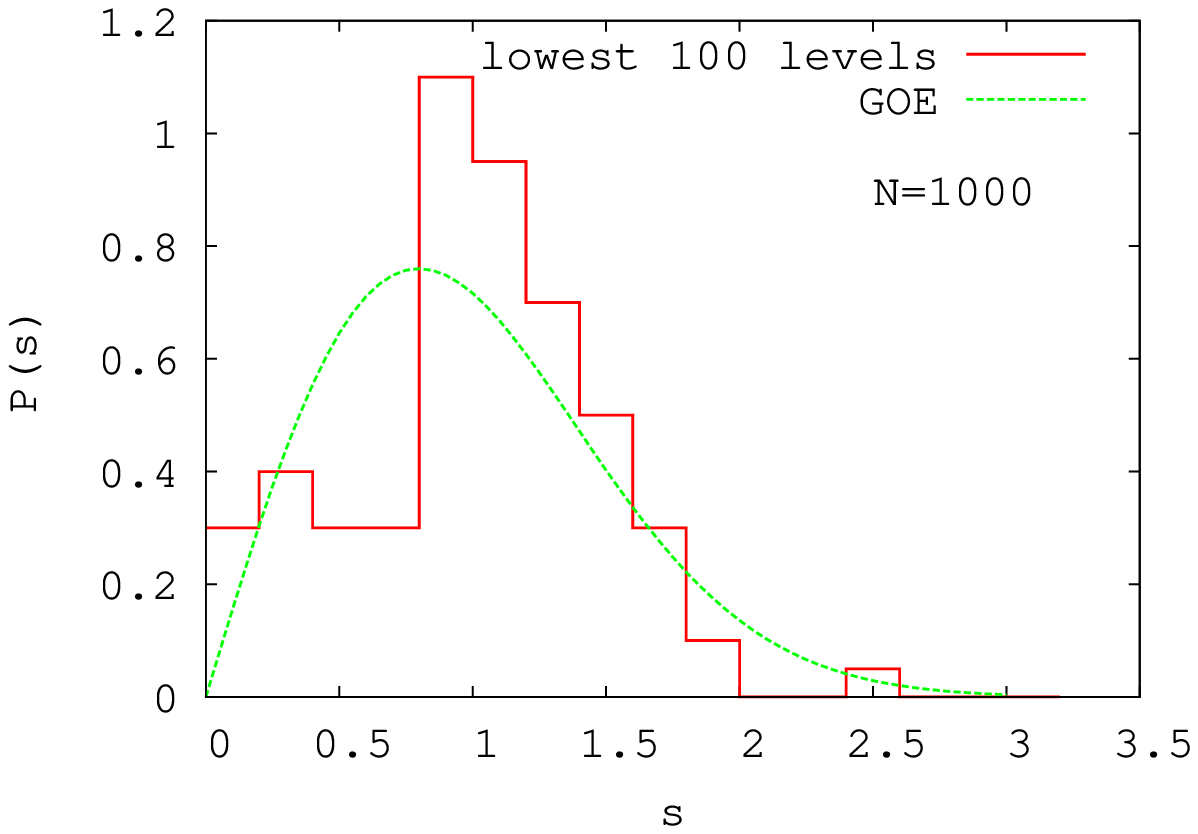}} \\
(a) \\
      \resizebox{65mm}{!}{\includegraphics[angle=0]{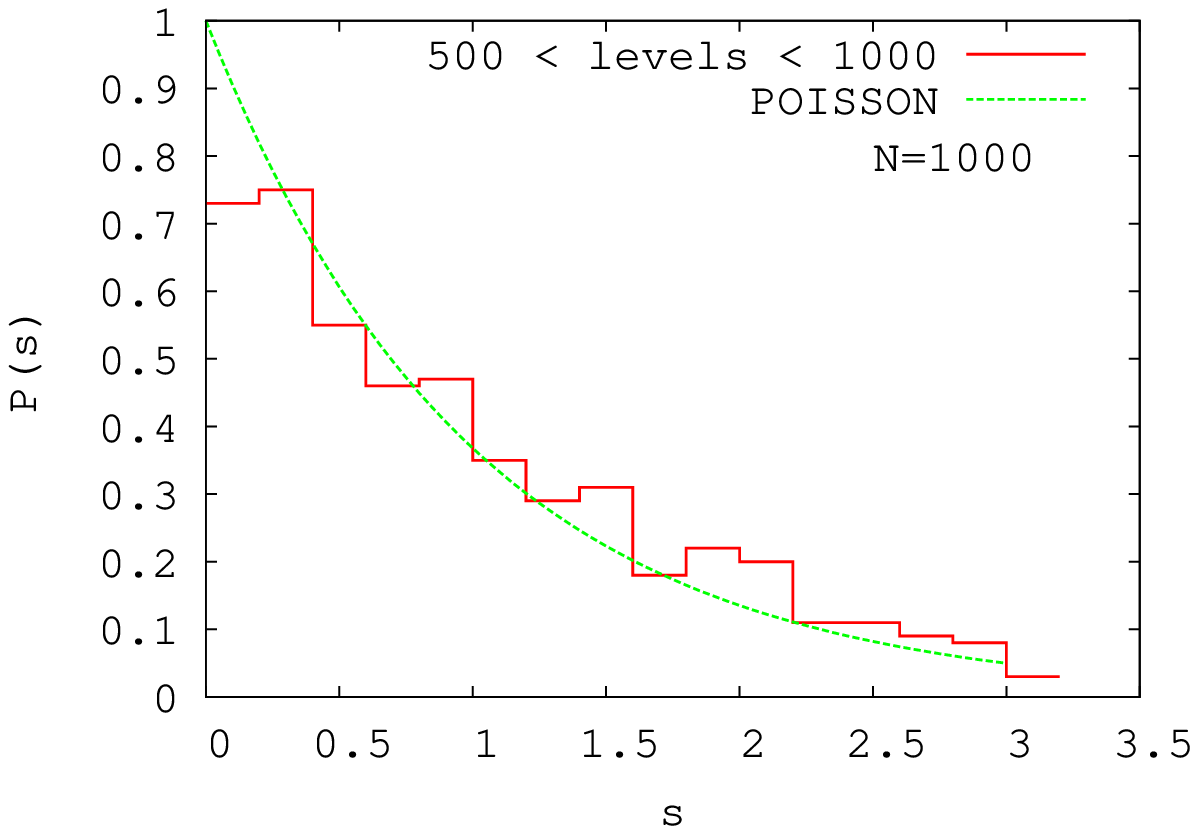}} \\
(b) \\
    \end{tabular}
  \end{center}
\caption{(Color online) Plot of the $P(s)$ distribution for different number of levels in different region of spectrum with $N = 1000$ $^{7}$Li atoms. In each panel the histograms present the $P(s)$ distribution obtained from many-body theory. The green dashed curve corresponds to the Wigner (GOE) distribution in panel (a) and in panel (b) it corresponds to the Poisson distribution.}
\end{figure}
\begin{figure}
  \begin{center}
    \begin{tabular}{cc}
      \resizebox{65mm}{!}{\includegraphics[angle=0]{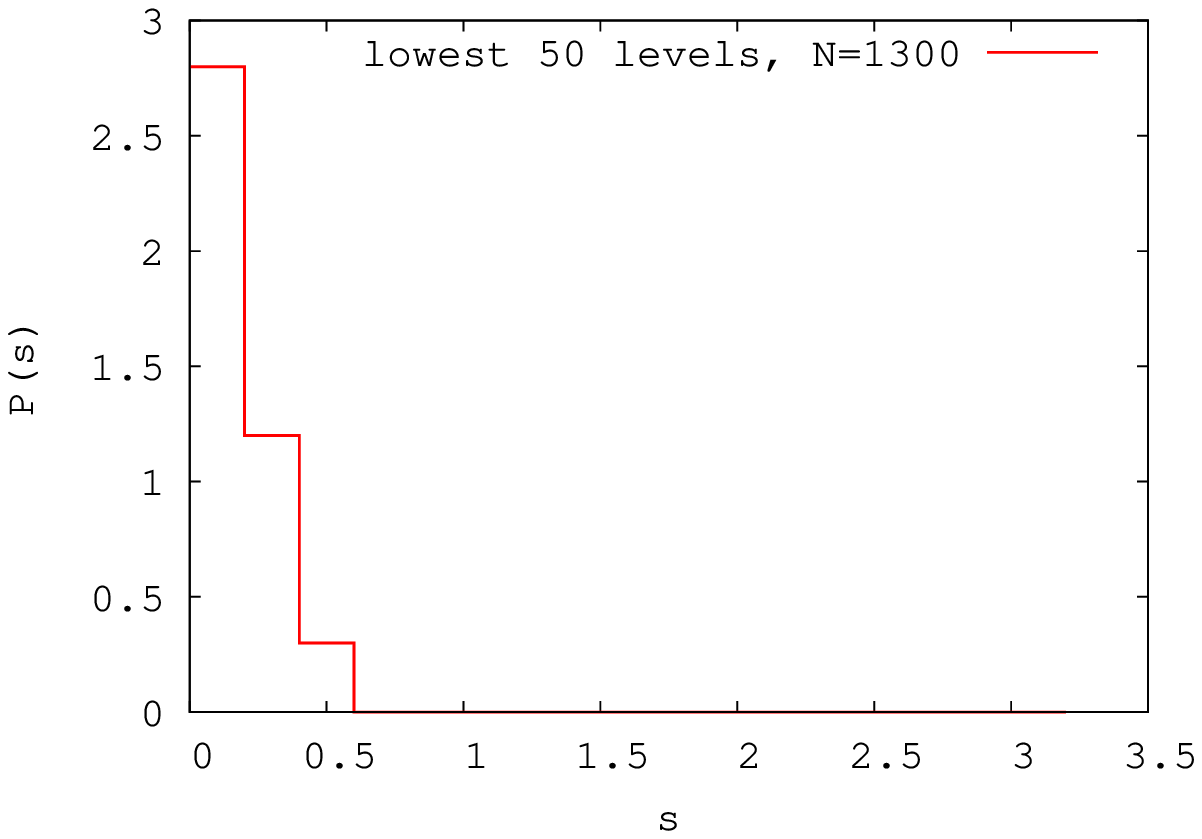}} \\
(a)\\
      \resizebox{65mm}{!}{\includegraphics[angle=0]{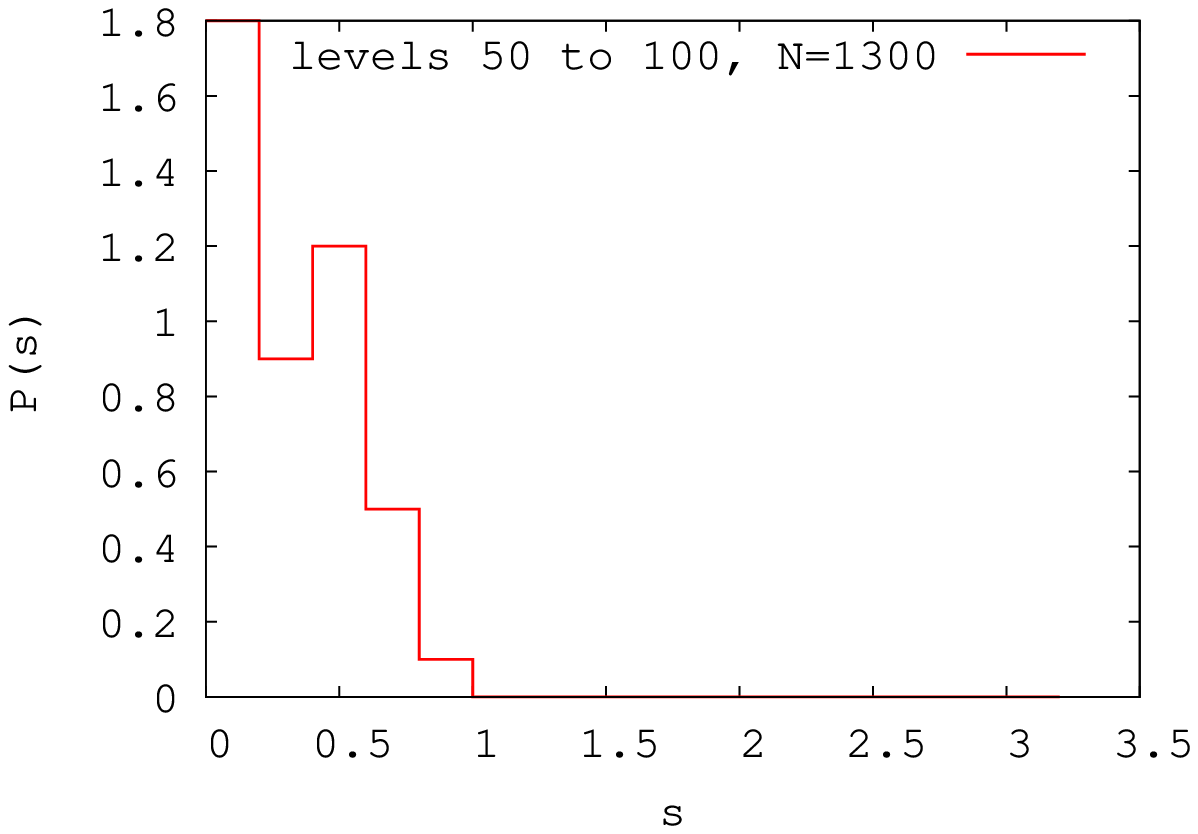}} \\
 
(b)\\
      \resizebox{65mm}{!}{\includegraphics[angle=0]{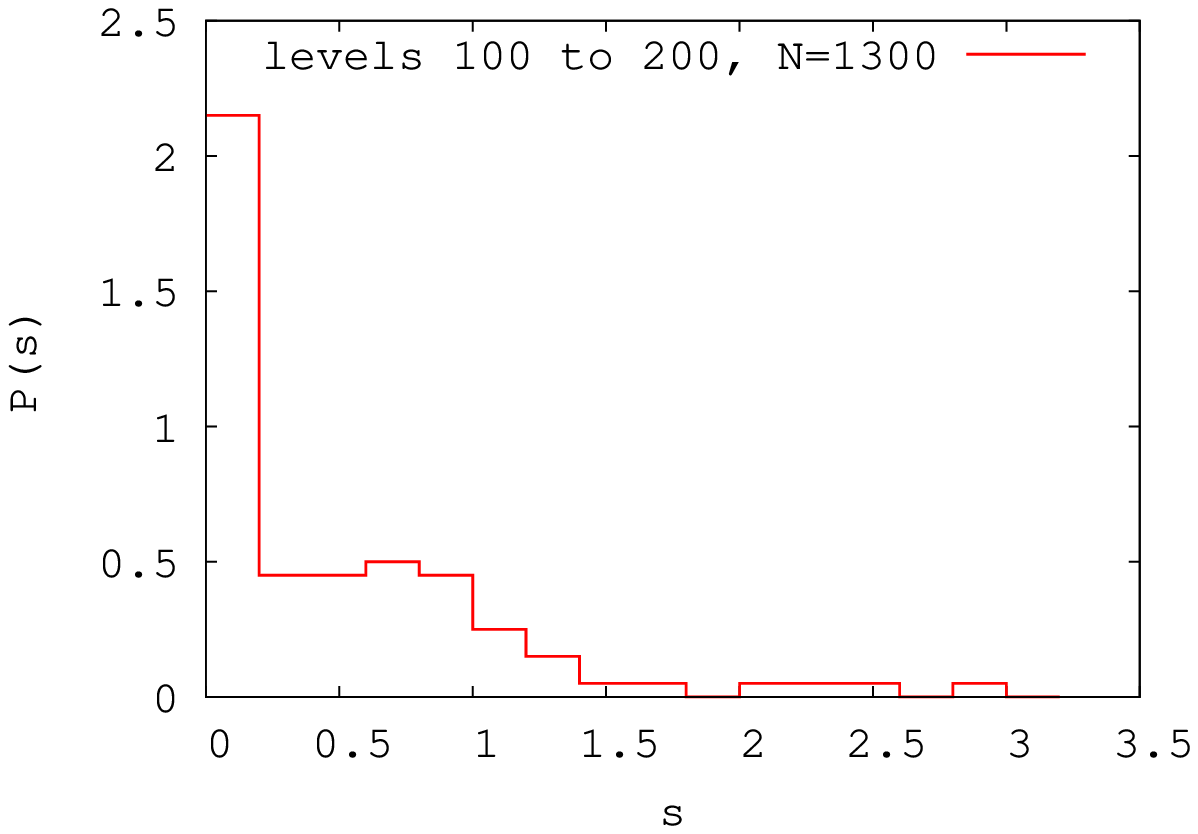}} \\
(c)\\
    \end{tabular}
  \end{center}
\caption{(Color online) The $P(s)$ distribution for different number of levels of $N=1300$ $^{7}$Li atoms in the trap is presented as histograms. Panel (a) corresponds to lowest 50 levels, panel (b) corresponds to lowest 50-100 levels and panel (c) corresponds to lowest 100-200 levels as indicated in each panel. 
}
\end{figure}
The situation becomes more interesting for $N=1300$ which is very close to the critical point and the condensate is highly correlated. It is reflected in Fig.~9(a)-(c) where we plot $P(s)$ for different levels. As near the critical point, the metastable region becomes flatter, we calculate lowest 200 levels for which the tunnelling probability through the intermediate barrier is negligible. For lowest 50 levels [Fig.~9(a)] we observe a sharp peak in the first bin near $s=0$. It signifies that many eigen-states overlap and it leads to the existence of large quasi-degenerate states. Although ${\mathcal V}(r)$ still dominates and the continuous symmetry of harmonic oscillator is removed, but discrete symmetry retains. In Fig.~9(b) and 9(c), although we get a sharp peak near $s=0$, but it spreads to further bins. It signifies that the effect of quasi-degeneracy is gradually lifted in the higher levels.\\  

\hspace*{.5cm}
   
\section{Concluding Remarks}
\hspace*{.5cm}
Using the correlated many-body technique we compute the energy spectrum of weakly interacting trapped Bose gas. All throughout our calculation 
the Bose gas is at zero temperature and under the harmonic confinement with fixed trap size which corresponds to the JILA experiment. Although the statistical behaviour of completely integrable and fully chaotic systems are understood, the intermediate region of integrability and chaos is more interesting. Interacting trapped bosons is such a system which is spatially inhomogeneous. The existence of external harmonic trap together with interatomic interaction makes the system more interesting. We study the spectral fluctuation and level correlation in the energy spectrum. Although there is no rigorous derivation, but the numerical results show a mixed statistics, which is very complexly dependent on the number of energy levels and number of the bosons in the trap. 
However for higher energy levels where the external trap is dominating we get back the Poisson type fluctuation, whereas the low-lying collective excitations are strongly influenced by interatomic interaction and shows level repulsion. Thus our findings do not strictly obey the earlier conjecture of Bohigas for atomic nucleus and atoms. The results for attractive Bose gas near the critical point is highly interesting which nicely shows how the degeneracy of the harmonic trap is gradually removed for higher levels. Although there is no experimental data for such high-lying states, but for dilute interacting Bose gas it is possible to measure experimentally with present-day set up. Our 
present study opens many questions for further study. In the present 
study the interacting Bos gas is in a fixed trap size. However the use of time-dependent potential will allow 
to study the dynamical behavior of energy spectrum and the time evolution of the spectral statistics and correlation properties. Throughout our calculation we use the zero-temperature Bose gas and delibarately avoids any thermal fluctuations. So it is also 
interesting to study the spectral distribution for non zero temperature BEC. Our present methodology is not 
valid as it avoids the thermal fluctuation.
\hspace*{.5cm}
\begin{center}
{\large{\bf{Acknowledgements}}}
\end{center}
This work is supported in part by DAE (Grant No. 2009/37/23/BRNS/1903), and DST (Fund No. SR/S2/CMP/0059(2007)). AB acknowledges CSIR for a senior reaserch fellowship (Grant No. 09/028(0773)-2010-EMR-1). SKH also acknowledges the Council of Scientific and Industrial Reaserch (CSIR), India for junior research fellowship (Grant No. 08/561(0001)/2010-EMR-1).


\end{document}